\documentclass[french]{gretsi}

\usepackage[utf8]{inputenc}

\usepackage[T1]{fontenc}

\usepackage[sorting = none]{biblatex}
\bibliography{biblio.bib}

\usepackage{subcaption}
\usepackage{csquotes}
\usepackage{algorithmic}
\usepackage{graphicx}
\usepackage{graphics}

\usepackage{amsmath, amssymb, amsfonts}

\begin{document}


\titre{Exploitation des propriétés de saturation synaptique pour obtenir un neurone à fréquence spécifique}

\auteurs{
  \auteur{Guillaume}{Marthe}{guillaume.marthe@insa-lyon.fr}{1}
  \auteur{Claire}{Goursaud}{claire.goursaud@insa-lyon.fr}{1}
}

\affils{
  \affil{1}{CITI, EA3720, 
Univ Lyon, INSA Lyon, Inria,
69621 Villeurbanne France \\
  }
}


\resume{Le principal frein au déploiement l'IoT reste la consommation énergétique des objets connectés. 
Plus précisément, les micro-contrôleurs consomment trop de puissance. Afin de remédier à ce problème, de nouveaux types de circuits ont été réalisés à partir de neurones à impulsions. Leurs calculs analogiques permettent une réduction drastique de la consommation.
Cependant, le domaine de l'analogique rend difficile les opérations séquentielles sur des signaux entrants, nécessaires dans un grand nombre d'applications.
Dans cet article, nous proposons d'utiliser un phénomène bio-inspiré appelé Interaction Synaptique afin de produire un filtre temporel. Nous proposons un modèle de synapses qui font que le neurone décharge pour une certaine plage de délais entre deux impulsions reçues, mais ne réagit pas si cet Inter-Spike Timing n'est pas compris dans la plage. Nous étudions les paramètres de ce modèle afin de comprendre comment adapter cette plage d'Inter-Spike Timings. L'originalité de ce papier est de proposer un nouveau moyen d'utiliser des séquences temporelles dans le domaine de l'analogique.}

\abstract{Energy consumption remains the main limiting factors in many promising IoT applications.
In particular, micro-controllers consume far too much power. In order to overcome this problem, new circuit designs have been proposed and the use of spiking neurons and analog computing has emerged as it allows a very significant consumption reduction. 
However, working in the analog domain brings difficulty to handle the sequential processing of incoming signals as is needed in many use cases.
In this paper, we propose to use a bio-inspired phenomenon called Interacting Synapses to produce a time filter. We propose a model of synapses that makes the neuron fire for a specific range of delays between two incoming spikes, but not react when this Inter-Spike Timing is not in that range. We study the parameters of the model to understand how to adapt the Inter-Spike Timing. The originality of the paper is to propose a new way, in the analog domain, to deal with temporal sequences.  }

\maketitle

\section{Introduction}

L'internet des objets (IoT) a permis le développement d'un grand nombre d'applications pour le quotidien \cite{InternetOT}, mais leur consommation énergétique reste un problème critique. 
En effet, la durée de vie des noeuds est prévue pour être de plusieurs années, mais cela ne peut être atteint par les batteries si le noeud est toujours actif. 
Cependant, la majorité des objets de l'IoT a besoin d'un envoie de données seulement de manière sporadique. 
Ainsi, une des solutions mise en place pour limiter leur consommation est de leur permettre de s'endormir, et de ne se réveiller que lorsqu'ils avaient besoin d'interagir. Mais, il est important de conserver la capacité de s'adresser à un objet en particulier, à n'importe quel moment, sans dépendre d'une synchronisation du système complet. 
Cela peut être réalisé à l'aide d'un récepteur Wake-Up Radio (WuR), dont le rôle est de surveiller le canal en continu et de réveiller le récepteur principal seulement si un signal spécifique est reçu \cite{WUR}. 

Jusqu'à présent, ces WuRs étaient réalisées à partir de micro-contrôleurs capable de réaliser de la reconnaissance de signature mais consommant des puissances autour de $200 \mu W$ \cite{perfWUR}. Cela apporte un gain par rapport à la consommation du récepteur principal, mais reste trop élevé pour permettre à la batterie de durer aussi longtemps que la durée de vie initialement prévue pour le noeud IoT \cite{lifetime}.

Les réseaux de neurones à impulsions (SNN) sont une technologie très basse consommation qui émerge dans le domaine du traitement du signal. Ceux-ci sont principalement utilisés pour de la reconnaissance d'image, ou pour l'étude de l'activité cérébrale en neurosciences. 
Ces systèmes bio-inspirés sont basés sur des transistors et reproduisent le comportement d'un neurone. Ils sont capables de calculer aussi vite que les machines actuelles, tout en consommant moins \cite{4fJ}.

Pour réduire encore plus la consommation d'énergie, nous considérons dans ce travail des circuits purement analogiques qui reproduisent le comportement d'un neurone \cite{sub35pW}. Jusqu'ici ces neurones ont été utilisés pour des approches combinatoires, qui ne comprennent pas de séquences temporelles. \cite{Mart2303:Wake} 
Des applications séquentielles comme de la vision 
ou du traitement du signal étaient donc impossibles à réaliser. Dans ce papier nous montrons qu'il est quand même possible d'introduire cette dépendance temporelle. Pour cela, nous proposons d'exploiter le phénomène biologique appelé interaction synaptique, et son effet mémoire \cite{articleRomain}. 
Nous exploitons ce phénomène afin que notre neurone réagisse à un rythme temporel particulier et évaluons l'impact des paramètres du neurone sur la sélectivité.

Ce papier est organisé comme suit : la section \ref{Models} présente le modèle du neurone et du système, ainsi que la manière dont ils ont été simulés. La section \ref{prop} décrit le comportement et l'implémentation du système. 
Le choix des paramètres et leur impact sont étudiés en section \ref{param}. La section \ref{ccl} conclue le papier.

\section{Modèles}\label{Models}
\subsection{Neurone LIF avec synapses classiques}

Un neurone biologique reçoit des impulsions des synapses entrantes, les agrège dans sa tension membranaire, puis envoie une impulsion au suivant lorsque cette tension a atteint un seuil donné. Un neurone bio-inspiré est basé sur le même principe. Il est modélisé par sa réponse à un train d'impulsions électriques. Sa tension interne varie en fonction du courant d'entrée. Il existe de nombreux modèles pour représenter la relation entre le courant d'entrée et la tension. Le modèle Leaky Integrate and Fire (LIF) est la référence, largement utilisé dans la littérature et est exprimé par :

\begin{equation}\label{eq1 : LIF}
    C_m \frac{dv}{dt} = - g_L(v(t) - v_{rest}) + I(t)
\end{equation}

avec $C_m$ la capacité de membrane, $v(t)$ le potentiel de membrane au temps $t$, $v_{rest} = -65mV$ le potentiel membranaire de repos, et $g_L$ la conductance de membrane. $I(t)$ modélise le courant synaptique d'entrée au cours du temps. Celui-ci peut être écrit, dans le cas où le signal d'entrée est une séquence d'impulsions, comme :

\begin{equation}\label{eq2 : spikeTrain}
   I(t) = \sum_{i} \delta (t - t_i)
\end{equation}

avec $t_i$ correspondant aux différents temps d'envoi d'impulsion, et $\delta$ à la fonction de Dirac.

A chaque fois que la tension atteint le seuil $v_{th}$, une impulsion est envoyée au neurone suivant et la tension est réinitialisée à $v_{rest}$. La courbe grise de la Fig. \ref{fig:4casesSSLIF}a représente l'évolution temporelle de la tension lors de la réception d'une impulsion unique. Dans un neurone LIF, on observe une augmentation instantanée de la tension lors de la réception d'une impulsion, puis une décroissance exponentielle due aux fuites de tension.

\subsection{LIF saturant}

Le modèle LIF est un modèle utile mais simplifié. En effet, il existe un autre phénomène qui est habituellement négligé mais que nous allons ici exploiter : la saturation des synapses. 
Cela modélise le fait que les paramètres de la synapse sont temporairement modifiés après la réception d'une impulsion. Dans ce cas, le courant interne ne dépend plus que du courant d'entrée, mais également de ce qu'il s'est passé précédemment. Ainsi, dans le modèle SLIF (LIF saturant), $I(t)$ (\ref{eq2 : spikeTrain}) est ici remplacé par $I_s(t)$ tel que :

\begin{equation}\label{eq3 : Is}
    I_s(t) = g_s(t) (E_s-v(t)) 
\end{equation}

\begin{figure}[!t]
\centering
\includegraphics[scale=0.4]{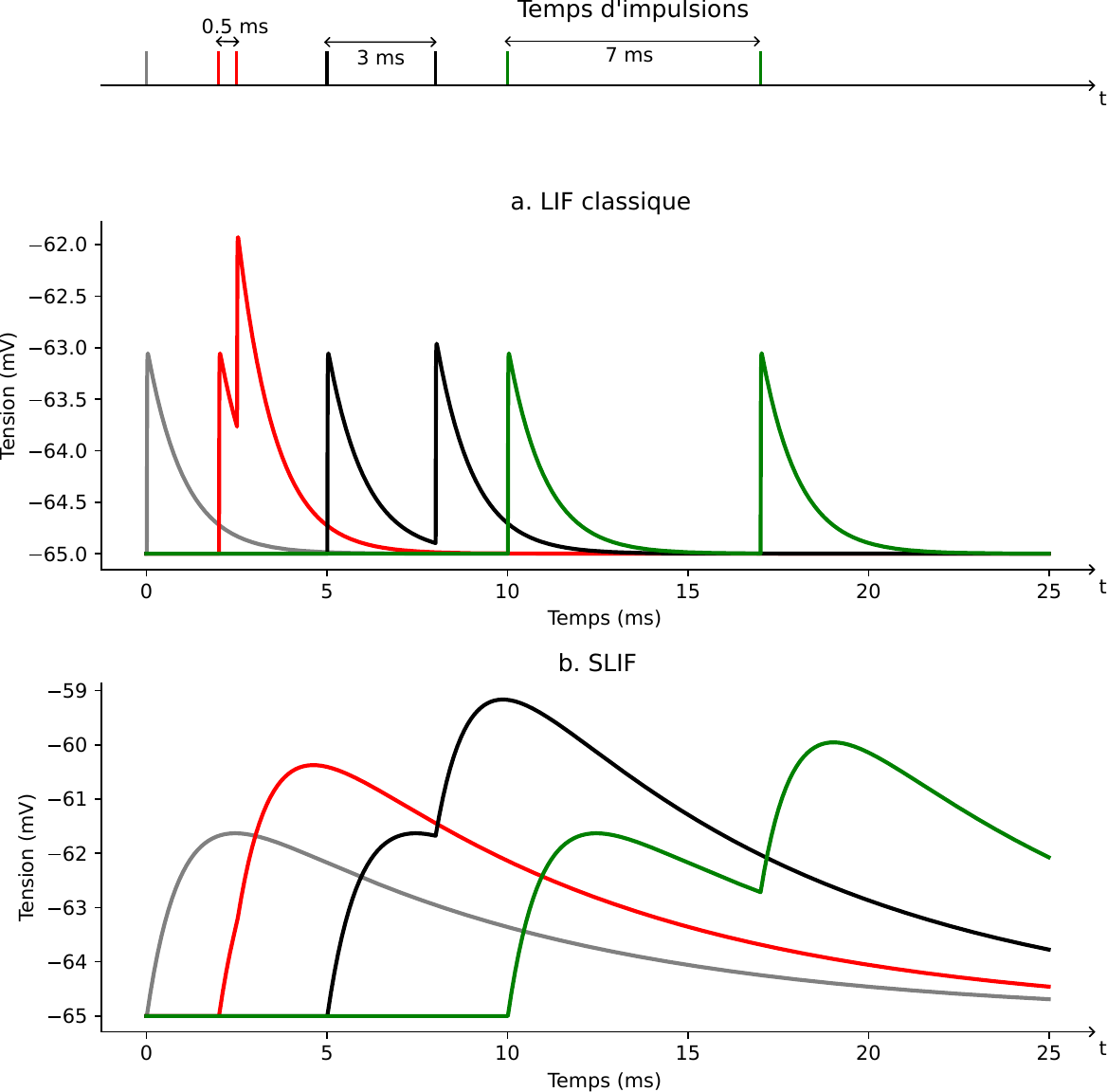}
\caption{Réponse du neurone à différents ISTs}
\label{fig:4casesSSLIF}
\end{figure}

Le courant synaptique dépend de la différence entre $v(t)$ et $Es = 0 mV$ le potentiel d'inversion synaptique. $g_s(t)$ est la conductance synaptique, variant au cours du temps et dépendant des séquences d'impulsions précédemment reçues. Cette variable est bornée entre 0 et sa valeur de saturation, fixée à $100pS$ pour nos simulations. Chaque fois qu'une synapse reçoit une impulsion, $g_s$ va augmenter jusqu'à sa borne supérieure, puis décroître exponentiellement selon :

\begin{equation}\label{eq4 : dgs}
    \frac{dg_s(t)}{dt} = \frac{-g_s(t)}{\tau_s} 
\end{equation}

avec $\tau_s$ la constante de temps synaptique. 
Les conséquences de cette saturation sont que la tension est contrainte, et met plus de temps à atteindre la tension prévue comme on peut le voir sur la courbe grise en Fig. \ref{fig:4casesSSLIF}b par rapport à celle en \ref{fig:4casesSSLIF}a.
Nous observons que la tension augmente doucement, puis décroît jusqu'à la tension de repos. Nous allons montrer comment cela peut être utilisé pour créer une mémoire à court terme pour des opérations temporelles.


\section{Proposition}\label{prop}

Notre objectif est d'utiliser cette structure neuronale pour décharger uniquement lors de la réception d'une cadence spécifique dans les trains d'impulsions reçus. Cela revient à sélectionner une période inter-impulsion précise. Pour cela, nous cherchons à atteindre une tension maximale si deux impulsions arrivent avec un temps inter-impulsion (IST) dans la plage prédéfinie.

Pour illustrer ce phénomène, nous envoyons deux impulsions consécutives à deux types de neurones LIF et SLIF, espacés de différents ISTs, qui représentent le temps entre la réception de ces deux impulsions.
Les résultats sont présentés en Fig. \ref{fig:4casesSSLIF}a et \ref{fig:4casesSSLIF}b avec quatre courbes chacune. 
La plus à gauche (gris) a été obtenue en envoyant seulement une impulsion aux neurones. Les trois autres ont été obtenues en envoyant deux impulsions, avec un IST différent à chaque fois. Les trois ISTs sont, de gauche à droite : $0.5ms$, $3ms$ et $7ms$. Les positions des impulsions correspondantes sont reportées en haut pour référence.

Lorsqu'un neurone LIF classique reçoit une impulsion, son potentiel de membrane croît soudainement. Puis, son potentiel décroît dû au phénomène de fuite du modèle. Lorsqu'une seconde impulsion est reçue, le potentiel croît à nouveau de la même amplitude, instantanément. L'amplitude maximale est donc atteinte lorsque l'IST est minimal, car 
la tension a moins de temps pour décroître. A contrario, si la seconde impulsion est envoyée longtemps après la première, le potentiel a plus diminué, et la tension est plus basse que dans le premier cas. L'amplitude obtenue décroît lorsque l'IST augmente.

La réponse du SLIF dans les mêmes conditions est montrée en Fig. \ref{fig:4casesSSLIF}b. Avec l'ajout de synapses saturantes, si une seconde impulsion est reçue juste après la première (cf. courbe rouge), la synapse est toujours saturée et cette impulsion ne fait pas croître le potentiel de membrane autant que la première. Sur la deuxième courbe (noire), l'IST est optimal, la saturation synaptique ne limite pas l'impact de la seconde impulsion, et le phénomène de fuite n'a pas assez de temps pour trop diminuer la tension, l'amplitude sera donc optimale. Pour la dernière courbe (verte), les impulsions sont trop espacées et le phénomène de fuite va faire décroître trop vite le potentiel avant la réception de la deuxième impulsion, ce qui va réduire l'amplitude maximale atteinte par la membrane. Ainsi, contrairement au LIF, l'augmentation d'amplitude du potentiel membranaire après la réception d'une deuxième impulsion n'est pas constante, et dépend de l'IST.

\begin{figure}[!t]
\centering
\includegraphics[scale=0.4]{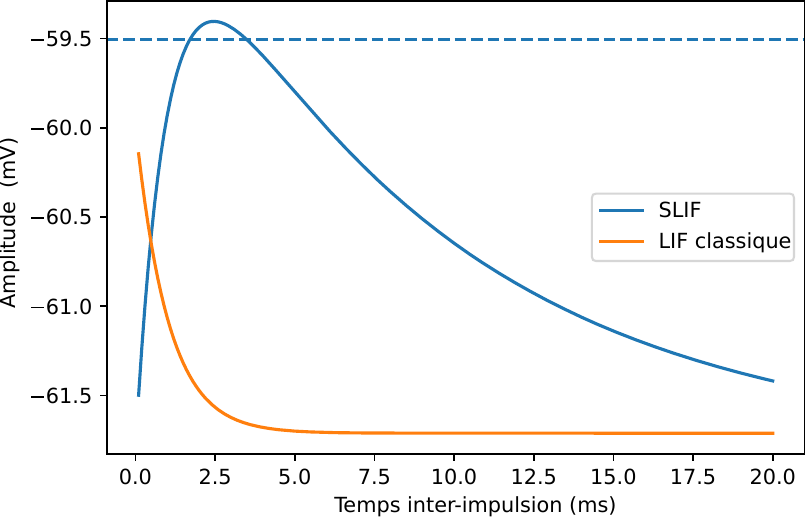}
\caption{Amplitude maximale atteinte par un SLIF et un LIF pour différents ISTs avec $C_m = 10^{-5} \mu F.cm^{-2}$, $g_L = 5 \cdot 10^{-5} S.cm^{-2}$ et $\tau_s = 0.9 ms$}
\label{fig:ampIST}
\end{figure}

\begin{figure}[!t]
\centering
\includegraphics[scale=0.4]{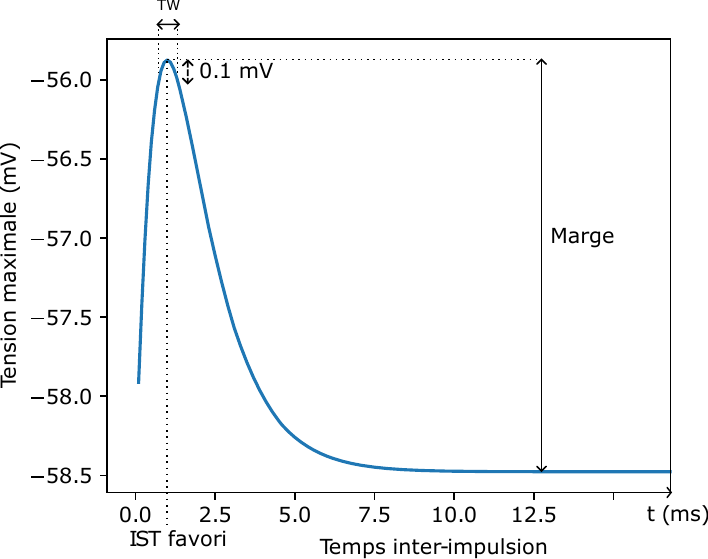}
\caption{Amplitude maximale atteinte par un SLIF avec $C_m = 10^{-3} \mu F.cm^{-2}$, $g_L = 10^{-5} S.cm^{-2}$ et $\tau_s = 0.01 ms$}
\label{fig:usrefAmp}
\end{figure}

Nous avons tracé les amplitudes atteintes en fonction de l'IST en Fig. \ref{fig:ampIST}. Pour le LIF, nous avons une décroissance exponentielle. Pour le SLIF, nous pouvons voir que l'amplitude maximale est atteinte pour un IST de 3ms et qu'il y a deux phases. Pour de faibles ISTs, la saturation entraîne une diminution de l'amplitude par rapport à l'optimum, avec une augmentation quasi linéaire jusqu'à l'IST favori puis une décroissance due aux fuites du neurone.

Le neurone décharge si son amplitude dépasse le seuil fixé (représenté par la ligne hachurée). 
Le seuil va pouvoir être utilisé pour discriminer une fine plage de valeurs d'IST. Dans cette simulation, le neurone décharge si les impulsions reçues étaient espacées de 2 à 3.5 ms. La largeur de cette plage d'ISTs sera appelée Timewidth (TW). 
Pour mesurer le TW, on impose un seuil 0.1 mV sous l'amplitude maximale, et on regarde la plage d'ISTs qui font que la tension membranaire atteint ce seuil.

L'autre métrique importante est la marge qui correspond à la différence entre les amplitudes maximale et minimale atteintes par la tension pour les ISTs considérés. L'objectif est d'optimiser ce jeu de paramètres. On veut obtenir la plus grande marge (pour être plus robuste aux perturbations), tout en réduisant le TW (pour être plus précis sur les ISTs sélectionnés). L'IST favori à viser dépend de la fréquence de motif choisie.

Pour cela on peut jouer sur les paramètres des équations :

\begin{itemize}
    \item 
$C_m$, la capacité de membrane qui permet de choisir les ordres de grandeur de la dynamique du neurone. 
    \item 
$g_L$, la conductance du neurone qui régule la force du phénomène de fuite de la tension membranaire. 
    \item 
$g_S$, la conductance synaptique. Sa valeur est impactée par sa constante de temps $\tau_S$ et modifie la vitesse d'évolution du potentiel de membrane lorsqu'une impulsion est reçue. 
\end{itemize}



\begin{figure*}[!t]
\centering
\begin{subfigure}{0.3\textwidth}
    \includegraphics[width=\columnwidth]{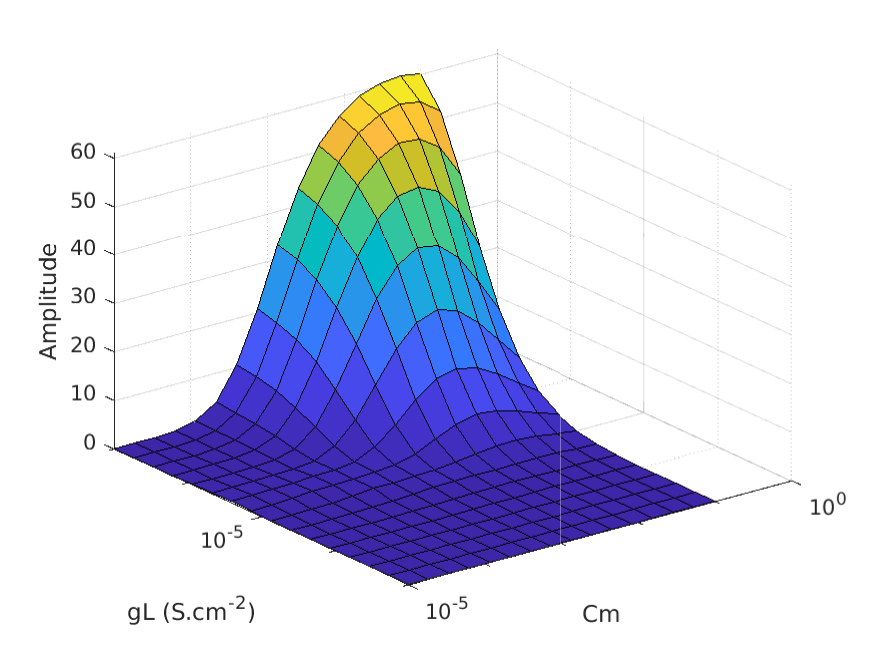}
    \caption{Marge maximale pour l'IST favori et $C_m \cdot\tau_s = 10^{-3} $}
    \label{fig:cmgLAmp}
\end{subfigure}
\hfill
\begin{subfigure}{0.3\textwidth}
    \includegraphics[width=\textwidth]{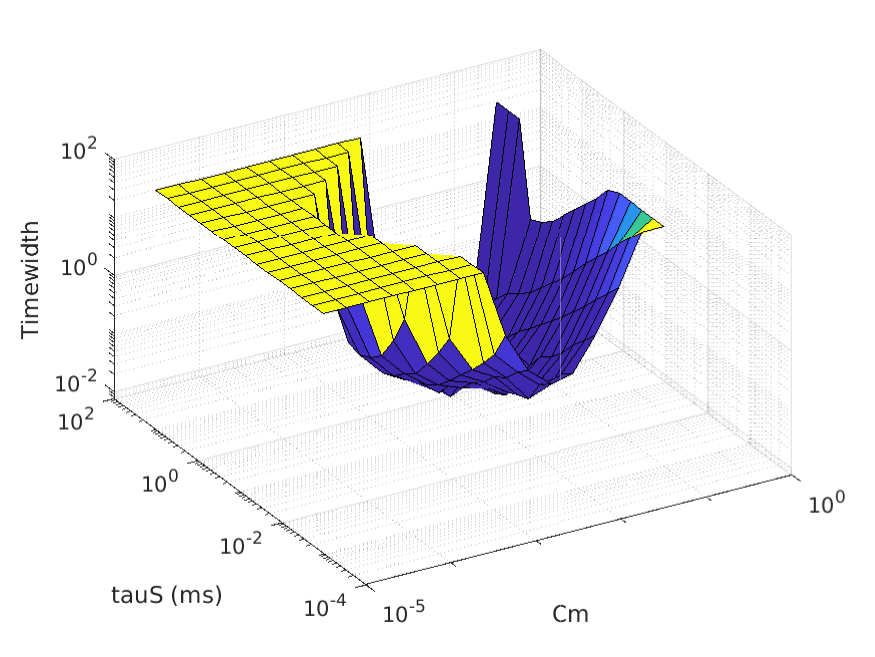}
    \caption{TW centré sur l'IST favori avec $g_L \cdot \tau_s = 10^{-5}$}
    \label{fig:cmtauSTW}
\end{subfigure}
\hfill
\begin{subfigure}{0.3\textwidth}
    \includegraphics[width=\textwidth]{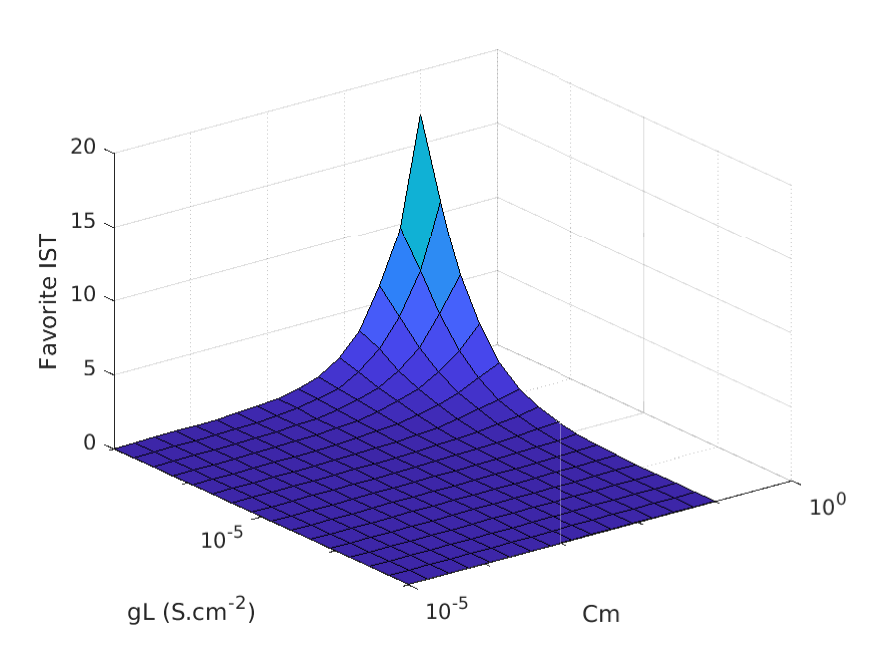}
    \caption{IST favori pour $C_m \cdot \tau_s = 10^{-3} $}
    \label{fig:CmgLIST}
\end{subfigure}
\vfill
\begin{subfigure}{0.3\textwidth}
    \includegraphics[width=\textwidth]{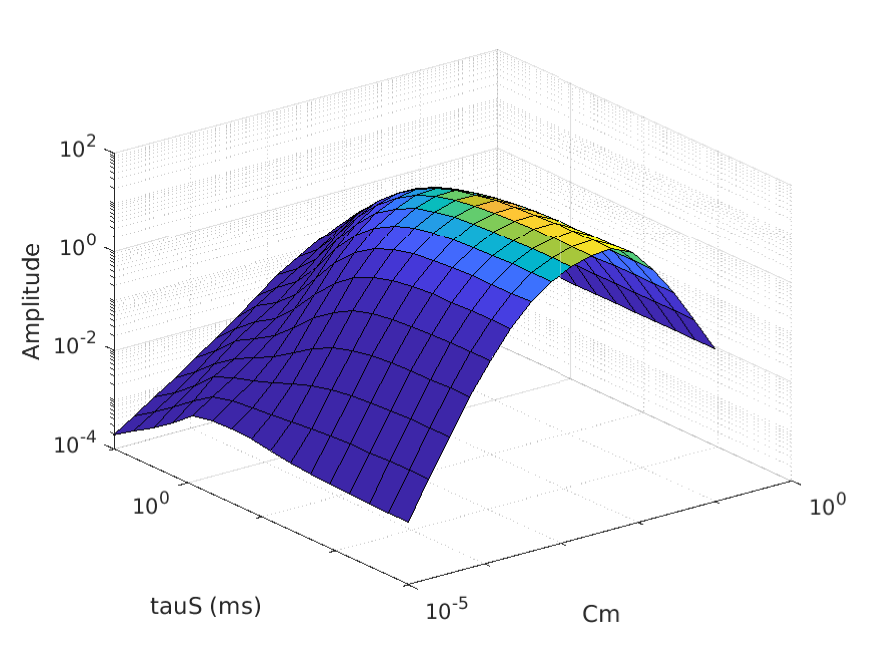}
    \caption{Marge maximale pour l'IST favori et $\tau_s \cdot g_L = 10^{-5} $}
    \label{fig:cmtausAmp}
\end{subfigure}
\hfill
\begin{subfigure}{0.3\textwidth}
    \includegraphics[width=\textwidth]{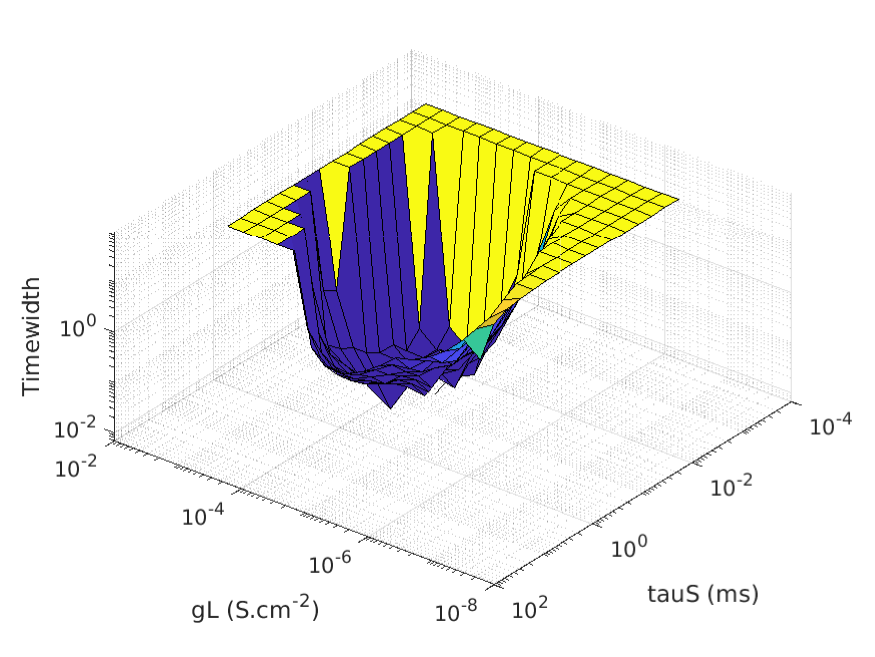}
    \caption{TW centré sur l'IST favori avec $g_L \cdot C_m = 10^{-7}$}
    \label{fig:tauSgLTW}
\end{subfigure}
\hfill
\begin{subfigure}{0.3\textwidth}
    \includegraphics[width=\textwidth]{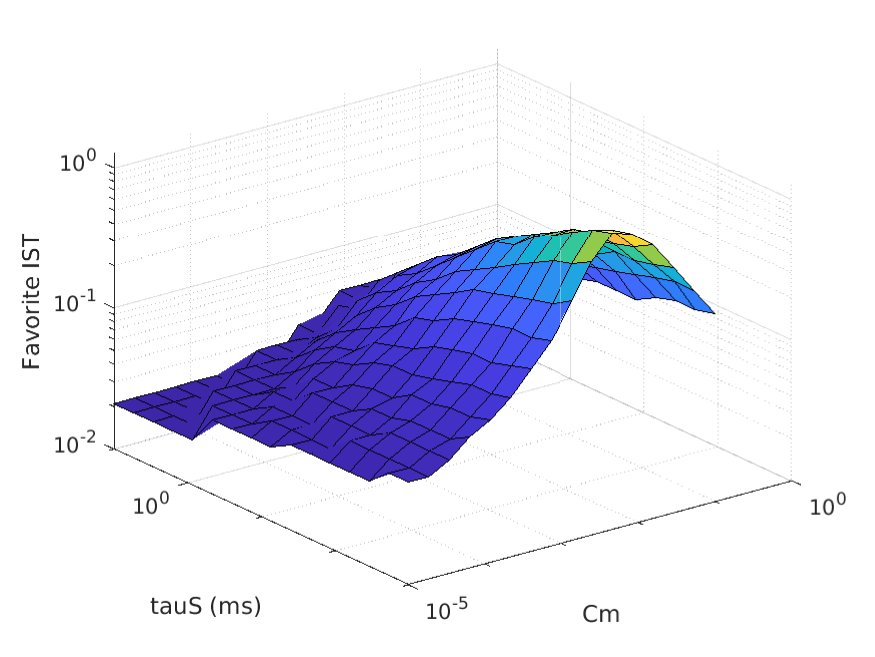}
    \caption{IST favori pour $\tau_s \cdot g_L = 10^{-5}$}
    \label{fig:CmtauSIST}
\end{subfigure}
\caption{Evolution de l'amplitude, du TW et de l'IST favori en fonction de $C_m$, $g_L$ et $\tau_s$}
\label{fig:figures}
\end{figure*}

\section{Étude paramétrique}\label{param}

Nous avons vu que le SLIF est un modèle qui permet de ne décharger que pour une plage spécifique d'ISTs, dépendant de nos paramètres. Nous étudions donc ici l'impact de ces paramètres sur la marge, le TW et l'IST favori. Pour ces études, nous avons choisi de fixer des valeurs de couples de paramètres à nos valeurs de référence, et d'observer les résultats en faisant varier à la fois un des paramètres du couple, et le paramètre libre. Nos couples de référence sont $C_m \cdot \tau_s = 10^{-3}$, $g_L \cdot \tau_s = 10^{-5}$ et $C_m \cdot g_L = 10^{-7}$. 

\subsection{Capacité de membrane \texorpdfstring{$C_m$}{h}}

$C_m$ impacte directement la dynamique du neurone, c'est pourquoi nous l'étudions en premier, afin de déterminer l'ordre de grandeur de l'IST favori. 
La capacité de membrane est directement multipliée à la dynamique du potentiel de membrane $\frac{dv}{dt}$. Nous pouvons voir en Fig. \ref{fig:cmgLAmp} lorsque $C_m$ augmente, que la marge maximale du potentiel de membrane augmente. Pour la Fig. \ref{fig:cmtausAmp}, on peut voir que l'on atteint une marge maximale pour une valeur de $C_m$ avoisinant $5 \cdot 10^{-3}$. On peut voir en Fig. \ref{fig:cmtauSTW} et \ref{fig:tauSgLTW} qu'on atteint un TW optimal, donc minimal, pour une valeur de $C_m$ avoisinant toujours $5 \cdot 10^{-3}$, et que le TW augmente si l'on s'éloigne de cette valeur. En Fig. \ref{fig:CmgLIST} et \ref{fig:CmtauSIST} on peut observer deux comportements différents pour l'évolution de l'IST en fonction de $C_m$. 
Lorsque le produit $C_m \cdot \tau_s$ est constant, la croissance de $C_m$ entraîne la croissance de la valeur d'IST favori. On voit que lorsque $\tau_s$ est constant, on atteint une valeur maximale de cet IST pour une valeur de $C_m$ de $5 \cdot 10^{-3}$. L'impact de $\tau_s$ sur l'IST est donc moins important que celui de $C_m$, mais plus que celui de $g_L$.

\subsection{Conductance de membrane \texorpdfstring{$g_L$}{h}}

$g_L$ est la conductance de membrane, qui définit la dynamique de la fuite de potentiel. Elle est directement proportionnelle à la décroissance exponentielle qui ramène à $v_{rest}$. Nous pouvons voir en Fig. \ref{fig:cmgLAmp} et \ref{fig:cmtausAmp} que lorsque $g_L$ augmente, la marge diminue, du fait de la fuite. Dans la même logique, on peut voir en Fig. \ref{fig:cmtauSTW}, \ref{fig:CmgLIST} et \ref{fig:CmtauSIST} que l'augmentation de $g_L$ entraîne une augmentation du TW et une réduction de l'IST favori.

\subsection{Constante de temps \texorpdfstring{$\tau_s$}{h}}

La valeur de $g_L$ est utilisée pour choisir la marge maximale que l'on souhaite atteindre. 
$\tau_s$ est la constante de temps de $g_s$, la conductance synaptique. Après la réception d'une impulsion, $g_s$ croît instantanément jusqu'à atteindre sa borne supérieure. Il va ensuite décroître jusqu'à sa borne inférieure. La vitesse de décroissance est régie par $\tau_s$. $g_s$ peut être associé à un poids dans notre modèle. Lorsque $\tau_s$ a une valeur basse, $g_s$ varie plus vite et plus fort. Cela améliore l'effet de la seconde impulsion sur le potentiel de membrane, car $g_s$ sature moins que pour une plus grande valeur de $\tau_s$ comme on peut le voir en Fig. \ref{fig:cmgLAmp} et \ref{fig:cmtausAmp}. 
$\tau_s$ va donc influencer TW. Lorsqu'il augmente, $g_s$ varie moins, donc le potentiel diminue plus lentement, et le TW devient plus large. Il a moins d'impact que les autres paramètres sur la marge et l'IST, on choisit donc de définir sa valeur en dernier, pour choisir le TW sans affecter les autres métriques.

Enfin, il est intéressant de noter qu'il n'y a pas besoin de faire de compromis entre le choix de la marge et celui du TW, car améliorer l'un améliore également l'autre.

\section{Conclusion}\label{ccl}

Dans cet article nous avons étudié un nouveau modèle de neurone impulsionnel bio-inspiré, basé sur des synapses saturantes, appelé SLIF. Nous avons montré que notre neurone répondait différemment selon l'Inter-Spike Timing entre les impulsions reçues. À la différence du LIF avec des synapses classiques, le SLIF permet d'atteindre une amplitude de tension membranaire maximale pour un IST configurable. Nous avons exploité ce phénomène naturel pour transformer un neurone impulsionnel et ses synapses en un filtre temporel qui émet une impulsion pour un IST choisi. Nous avons montré qu'il faut d'abord choisir $C_m$ puis $g_L$, puis $\tau_s$ pour paramétrer le système comme souhaité. Nos futurs travaux seront orientés sur l'analyse des SLIFs pour la discrimination de trains d'impulsions basée sur la cadence d'impulsion, afin de réaliser une Wake-Up Radio basse puissance.

\printbibliography

@ARTICLE{InternetOT,
  title={Internet of Things: Vision, Applications and Challenges},
  author={Rishika Mehta and Jyoti Sahni and Kavita Khanna},
  journal={Procedia Computer Science},
  year={2018},
  volume={132},
  pages={1263-1269}
}

@ARTICLE{WUR,
  author={Demirkol, Ilker and Ersoy, Cem and Onur, Ertan},
  journal={IEEE Wireless Communications}, 
  title={Wake-up receivers for wireless sensor networks: benefits and challenges}, 
  year={2009},
  volume={16},
  number={4},
  %pages={88-96},
  %doi={10.1109/MWC.2009.5281260}
}

@ARTICLE{4fJ,
  
AUTHOR={Sourikopoulos, Ilias and Hedayat, Sara and Loyez, Christophe and Danneville, François and Hoel, Virginie and Mercier, Eric and Cappy, Alain},   
	 
TITLE={A 4-fJ/Spike Artificial Neuron in 65 nm CMOS Technology},      
	
JOURNAL={Frontiers in Neuroscience},      
	
VOLUME={11},      
	
YEAR={2017},      
	  
%URL={https://www.frontiersin.org/article/10.3389/fnins.2017.00123},       
	
%DOI={10.3389/fnins.2017.00123},      
	
%ISSN={1662-453X},   
   
ABSTRACT={As Moore's law reaches its end, traditional computing technology based on the Von Neumann architecture is facing fundamental limits. Among them is poor energy efficiency. This situation motivates the investigation of different processing information paradigms, such as the use of spiking neural networks (SNNs), which also introduce cognitive characteristics. As applications at very high scale are addressed, the energy dissipation needs to be minimized. This effort starts from the neuron cell. In this context, this paper presents the design of an original artificial neuron, in standard 65 nm CMOS technology with optimized energy efficiency. The neuron circuit response is designed as an approximation of the Morris-Lecar theoretical model. In order to implement the non-linear gating variables, which control the ionic channel currents, transistors operating in deep subthreshold are employed. Two different circuit variants describing the neuron model equations have been developed. The first one features spike characteristics, which correlate well with a biological neuron model. The second one is a simplification of the first, designed to exhibit higher spiking frequencies, targeting large scale bio-inspired information processing applications. The most important feature of the fabricated circuits is the energy efficiency of a few femtojoules per spike, which improves prior state-of-the-art by two to three orders of magnitude. This performance is achieved by minimizing two key parameters: the supply voltage and the related membrane capacitance. Meanwhile, the obtained standby power at a resting output does not exceed tens of picowatts. The two variants were sized to 200 and 35 μm<sup>2</sup> with the latter reaching a spiking output frequency of 26 kHz. This performance level could address various contexts, such as highly integrated neuro-processors for robotics, neuroscience or medical applications.}
}

@article{sub35pW,
title = {A Sub-35 pW Axon-Hillock artificial neuron circuit},
journal = {Solid-State Electronics},
volume = {153},
pages = {88-92},
year = {2019},
%issn = {0038-1101},
%doi = {https://doi.org/10.1016/j.sse.2019.01.002},
%url = {https://www.sciencedirect.com/science/article/pii/S0038110118305562},
author = {F. Danneville and C. Loyez and K. Carpentier and I. Sourikopoulos and E. Mercier and A. Cappy},
keywords = {CMOS, Axon-Hillock, Artificial neuron, Ultra low power, High energy efficiency},
abstract = {Artificial Intelligence (AI) applications are developing at a high rate, facing soon a tremendous energy challenge. In this context, the original Axon-Hillock (AH) Artificial Neuron (AN) has been optimized to achieve ultra-low power (ULP) consumption. The membrane capacitance was taken out, and in order to drastically reduce its power consumption, the (feedback) capacitance is lowered to 5 fF, the transistors gate width is reduced to 120 nm and the supply voltage is decreased to as low as 200 mV. Designed and fabricated using 65 nm CMOS Technology, the refined AH neuron features a standby power of 11 pW, and when excited, a power consumption that does not exceed 30 pW for a firing frequency of 15.6 kHz. Its energy efficiency per spike is lower than 2 fJ/spike when the DC power is included (around 1 fJ/spike excluding the DC power), for an area of 31 µm2. These performance confer to this ULP AH neuron a high potential for future development of highly energy efficient Spiking Neural Networks, required to design future neuroprocessors embedded in various applications (smart visual sensors for autonomous vehicles, robotics).}
}

@ARTICLE{lifetime,
  author={Yetgin, Halil and Cheung, Kent Tsz Kan and El-Hajjar, Mohammed and Hanzo, Lajos Hanzo},
  journal={IEEE Communications Surveys \& Tutorials}, 
  title={A Survey of Network Lifetime Maximization Techniques in Wireless Sensor Networks}, 
  year={2017},
  volume={19},
  number={2},
  %pages={828-854},
  %doi={10.1109/COMST.2017.2650979}
}

@ARTICLE{perfWUR,
  author={Mazloum, Nafiseh Seyed and Edfors, Ove},
  journal={IEEE Transactions on Wireless Communications}, 
  title={Performance Analysis and Energy Optimization of Wake-Up Receiver Schemes for Wireless Low-Power Applications}, 
  year={2014},
  %volume={13},
  %number={12},
  %pages={7050-7061},
  %doi={10.1109/TWC.2014.2334658}
  }

@ARTICLE{articleRomain,
  author={Cazé RD},
  journal={F1000Research}, 
  title={All neurons can perform linearly non-separable computations}, 
  year={2022},
  %doi={https://doi.org/10.12688/f1000research.53961.3}
  }

@INPROCEEDINGS{Mart2303:Wake,
AUTHOR="Guillaume Marthe and Claire Goursaud and Laurent Clavier",
TITLE="Wake-up radio receiver based on Spiking Neurons for detecting activation
sequence",
BOOKTITLE="IEEE WCNC 2023",
%ADDRESS="Glasgow, United Kingdom (Great Britain)",
DAYS="26",
MONTH=mar,
YEAR=2023,
KEYWORDS="Spiking Neural Network; Wake-Up Radios; Active sequence detection",
ABSTRACT="Energy consumption is a critical issue for the deployed nodes in the area
of Internet of Things (IoT). This is the reason why many research focus on
Wake-up Radio (WuR) receivers that permit to let the nodes in sleep mode as
long as possible and to wake them up only if needed. However, current WuR
use classic microcontrollers that are still too energy consuming.
Meanwhile, Spiking Neural Networks (SNN) offer much lower power
consumption. Thus, we propose to adapt those neural networks as a wake-up
radio receiver in the IoT context. We aim at waking up the concerned node
by recognising one or many activation sequences in a bit flow. We propose
here a configuration for the neurons along with the design of appropriate
sequences. We show the performances of our system, and the the impact of
different parameters on the recognition accuracy."
}

\end{document}